# Imaging Earth-like planets with Extremely Large Telescopes

A. Chelli

Laboratoire d'Astrophysique de Grenoble and Jean-Marie Mariotti Center, F-38041 Grenoble Cedex 9
e-mail: `Alain.Chelli@obs.ujf-grenoble.fr`

**Abstract.** We investigate the possiblity to detect Earth-like planets, in the visible and the near infrared domains, with ground based Extremely Large Telescopes equipped with adaptive systems capable of providing high Strehl ratios. From a detailed analysis of the speckle noise, we derive analytical expressions of the signal to noise ratio on the planet flux, for direct and differential imaging, in the presence of the speckle noise and the photon noise of the residual stellar halo. We find that a 100m telescope would detect an Earth at a distance of 10pc, with a signal to noise ratio of 5, in an integration time of 12 hours. This requires to control the instrumental aberrations with a precision better than 1 nanometer rms, and to reach an image dynamics of $1.2 \times 10^6$ at $0.''1$ radius. Under the same conditions, a telescope of 30m would require a dynamics of $1.3 \times 10^7$ for a positive detection.

## 1. Introduction

Since 1995, more than 130 planets have been found through radial velocity measurements surveys (Udry et al. 2003). The next important step forward will be the detection and characterization of exoplanets with ground based Extremely Large Telescopes (Angel 2003). The aim of the present work is twofold: (1) establish analytic expressions of the signal to noise ratio on the planet flux in the presence of speckle and photon noise, (2) study the possible detection of an Earth-like planet, in the visible and the near infrared domains, with an Extremely Large Telescope. We compute the spatial intensity distribution of the long exposure coronagraphic image in Section 2. In Section 3, we introduce the spatial and spectral covariance coefficient of the speckle noise. Analytical expressions of the signal to noise ratio (SNR) on the planet flux are derived in Section 4 and Section 5, for direct imaging and differential imaging. Finally, the results for an Earth-like planet are discussed in Section 6.

## 2. Long exposure coronagraphic image

Let us consider an Extremely Large Telescope provided with a monolithic pupil of diameter $D$ and equipped with an adaptive optics system capable of providing high Strehl ratios ($> 0.5$). We observe in the focal plane of the telescope the image of an unresolved star through a $\pi$-rotation shearing coronagraph (Gay et al. 1997, Roddier & Roddier 1997), whose effect is to reduce the image dynamics by cancelling the coherent part of the light. The imaging system, telescope included, is assumed to be free of instrumental aberrations.

At high Strehl ratios $\mathcal{S}$, the long exposure image of the telescope at the angular position $\alpha$ and at the wavelength $\lambda$, partially corrected by adaptive optics, can be written as (see Appendix A):

$$< I_t(\alpha, \lambda) > = K \left[ \mathcal{S} \frac{\pi D^2}{4\lambda^2} A(\alpha) + (1 - \mathcal{S}) < L(\alpha, \lambda) > \right] \quad (1)$$

where $K$ is the average number of photoelectrons, $A(\alpha)$ is the Airy disc function normalized at the origin and $< L(\alpha, \lambda) >$ is the incoherent part of the image normalized in energy. The intensity at the ouput of the coronagraph is:

$$< I_c(\alpha, \lambda) > \approx K(1 - \mathcal{S}) < L(\alpha, \lambda) > [1 - A(2\alpha)] \quad (2)$$

It coincides with the incoherent part of the image plus a depression of half an Airy disk size in the center.

The incoherent part of the telescope point spread function, $(1 - \mathcal{S}) < L(\alpha, \lambda) >$, is equal to the power spectrum of the residual phase over the telescope pupil (Rigaut et al. 1998). In the simple analytic model proposed by Jolissaint & Veran (2002), the latter is the sum of five terms. We keep the two dominant terms, i.e. the servo-lag spectrum and the uncorrected high frequency spectrum, assuming a noise free wavefront sensor, no aliasing (Verinaud et al. 2004) and no anisoplanatism. Figure 1 shows a radial cut of the coronagraphic images in the V (0.55$\mu$m), R (0.70$\mu$m), I (0.90$\mu$m), J(1.25$\mu$m) and H (1.65$\mu$m) photometric bands, with an actuator spacing of 20cm, a seeing of $0.''8$ at $0.50\mu$m, an outer scale of the atmosphere of 25m, a wind speed of 10m/s and time constants of 0.37ms for

*Send offprint requests to*: A. Chelli

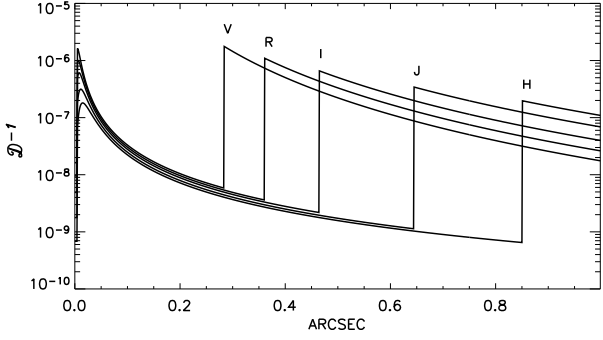

**Fig. 1.** Intensity at the output of the coronagraph at 0.55, 0.70, 0.90, 1.25 and 1.65$\mu m$, normalized to the maximum of the perfect point spread function. The curves represent the inverse of the image dynamics as defined in Section 2 ($D = 100m$; Fried parameter: $0.16(\lambda/0.50)^{1.2}$; actuators spacing: 20cm; outer scale of the atmosphere: 25m; wind speed: 10m/s; time constants: 0.37ms). The Strehl ratios are 0.67, 0.80, 0.88, 0.94 and 0.965, respectively

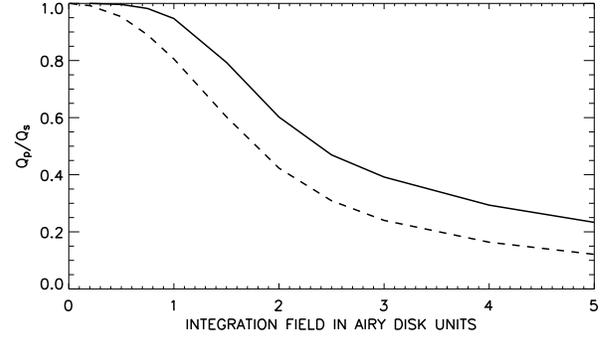

**Fig. 2.** Ratio $Q_p/Q_s$, normalized for a zero field, as a function of the integration field for monochromatic (full line) and broad band (broken line, $\lambda/\Delta\lambda = 5$) images. Note that this ratio is independent of the telescope diameter, and beyond a few Airy disks distance, its shape is also independent of $\gamma$.

the adaptive correction. The Strehl ratios are 0.67, 0.80, 0.88, 0.94 and 0.965, respectively.

To quantify the image quality, we introduce the dynamics of the image, $\mathcal{D}(\gamma) = \frac{\pi D^2}{4\lambda^2}(1-S)^{-1} <L(\gamma,\lambda)>^{-1}$, defined as the ratio of the intensity of the perfect stellar image in the center of the field to the intensity of the incoherent halo at the angular position $\gamma$. For a 100m telescope, the dynamics at the distance $\gamma = 0\farcs1$ varies from $3\times10^7$ at $0.55\mu m$ to $4.5\times10^7$ at $1.65\mu m$. As the dynamics is proportional to $D^2$, it is 10 times lower for a 30m telescope.

## 3. Spatial and spectral covariance of speckle noise

The speckle noise is related to the intensity variations produced by the phase fluctuactions over the telescope pupil. In practice, the signal, hence the speckle noise, are integrated over a field of view $\omega$ and a spectral bandwidth $\delta\lambda$. In order to make a proper derivation of the speckle noise, we need to compute its spatial and spectral covariance $C(\alpha,\lambda_1;\beta,\lambda_2)$. In the limit of high Strehl ratios, it is (see appendix B):

$$C(\alpha,\lambda_1;\beta,\lambda_2) \approx 2 <I(\alpha,\lambda_1)><I(\beta,\lambda_2)> A(\alpha l_1 - \beta l_2)$$
$$\approx 2 <I(\alpha,\lambda)>^2 l_1^4 l_2^4 S^{l_1^2+l_2^2-2} A(\alpha l_1 - \beta l_2) \quad (3)$$

where $l_1 = \lambda/\lambda_1$ and $l_2 = \lambda/\lambda_2$.

The variance of the speckle noise is equal to twice the square of the halo intensity. If we exclude the factor of two resulting from the addition of two incoherent speckle fields in the coronagraphic image, the same result has been established by Dainty (1974) for pure turbulent images and extended later to the incoherent part of partially corrected images (Goodman 1985, Canales & Cagigal 1999, Aime & Soummer 2004). Hence, Eq. (3) is valid at any Strehl ratios. The correlation coefficient of the speckle noise is equal to the Airy disk function $A(\alpha l_1 - \beta l_2)$.

## 4. Imaging

We concentrate on a system formed by an unresolved star in the center of the field and a planet located at a distance $\gamma$ larger than a few Airy disks. The image of the planet appears as 2 symmetrical contributions superposed on the stellar halo. Adding the two symmetrical contributions of the planet provides a useful signal $P = fSKQ_p$, where $f$ is the flux ratio between the planet and the star. $Q_p$ is given by:

$$Q_p = \frac{\pi D^2}{4\lambda^2}\frac{1}{\delta\lambda}\int_{\omega^2,\delta\lambda} l_1^2 A(\alpha l_1) d\alpha d\lambda_1 \quad (4)$$

Assuming a perfect detector, the limiting noises are the speckle noise $\sigma_s$ and the photon noise $\sigma_p$ from the stellar halo. Hence, the variance of $P$ is $\sigma^2(P) = 4\sigma_s^2 + 2\sigma_p^2$, with:

$$\sigma_s^2 = 2K^2(1-S)^2 <L(\gamma,\lambda)>^2 \left(\frac{4\lambda^2}{\pi D^2}\right)^2 Q_s^2(\overline{\lambda}_1,\overline{\lambda}_1)\frac{\tau}{T} \quad (5)$$

$$\sigma_p^2 = K(1-S)<L(\gamma,\lambda)>\omega^2 \quad (6)$$

and

$$Q_s^2(\overline{x},\overline{y}) = \left(\frac{\pi D^2}{4\lambda^2}\right)^2 \frac{1}{\delta\lambda^2}\int_{(\omega^2,\delta\lambda)^2} d\alpha d\beta dx dy$$
$$S^{(\frac{\lambda}{x})^2+(\frac{\lambda}{y})^2-2}\left(\frac{\overline{x}}{x}\right)^4\left(\frac{\overline{y}}{y}\right)^4 A\left[\left(\alpha\frac{\overline{x}}{x}-\beta\frac{\overline{y}}{y}\right)+\gamma\left(\frac{\overline{x}}{x}-\frac{\overline{y}}{y}\right)\right] \quad (7)$$

where $T$ is the total integration time and $\tau$ is the speckle lifetime. For bright sources, the speckle noise dominates and the SNR is independent of the photon flux. It is given by:

$$\text{SNR} = \frac{\pi f}{8\sqrt{2}}\frac{S}{1-S}\frac{1}{<L(\gamma,\lambda)>}\left(\frac{D}{\lambda}\right)^2\frac{Q_p}{Q_s}\sqrt{\frac{T}{\tau}} \quad (8)$$

Except for the ratio $Q_p/Q_s$ and the factor $\sqrt{2}$, the same expression has been obtained experimentally from pure speckle images by Racine et al (1999).

Figure 2 shows the variation of the ratio $Q_p/Q_s$ as a function of the integration field in Airy disk units, for monochromatic and low spectral resolution images ($\mathcal{R} = \lambda/\delta\lambda = 5$).

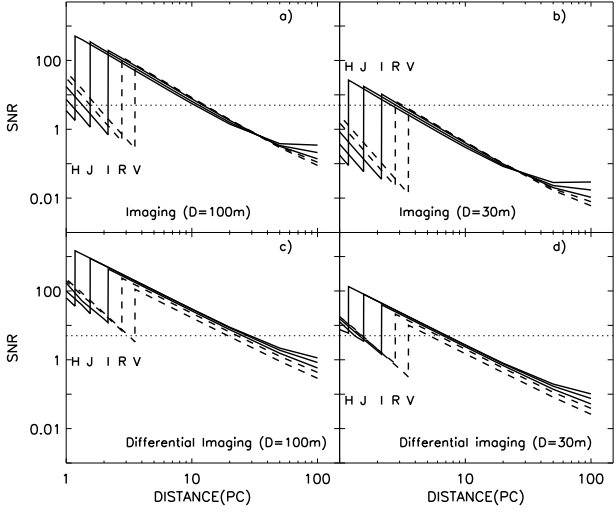
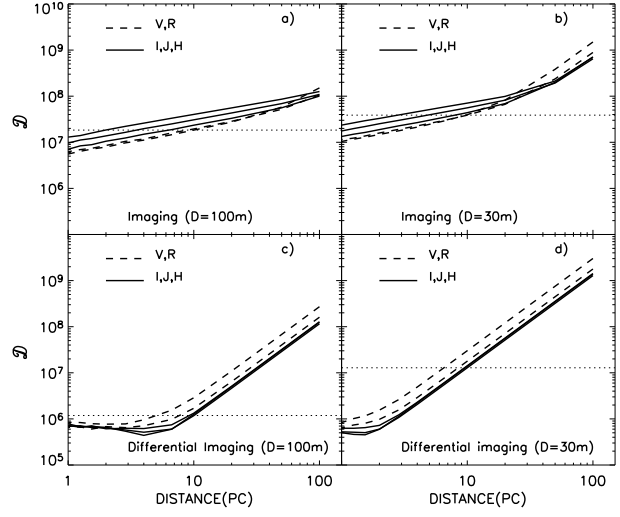

**Fig. 3.** Signal to noise ratios on the planet flux as a function of the distance, for imaging (a and b: $\lambda/\delta\lambda = 5$) and full contrast differential imaging (c and d: $\epsilon = 1$, a spectral resolution of 100 and the combined signal of 20 bands), for an integration time of 12 hours and telescope sizes of 30 and 100m, respectively (the dotted lines correspond to a SNR of 5). We adopt a constant Earth to Sun flux ratio of $1.5 \times 10^{-10}$ and Sun fluxes at 10pc of 45, 54, 55, 54, and 47Jy in the V, R, I, J and H photometric bands, respectively. For the calculations we used the coronagraphic images of Figure 1, a speckle lifetime of $0.01(\lambda/0.50)^{1.2}$, an effective aperture of $0.85D$, and an instrumental efficiency t=0.2. The SNR's at 10pc are 0.4 and 3 for a 30m telescope and, 8 and 30 for a 100m telescope, with imaging and differential imaging, respectively.

**Fig. 4.** Image dynamics required to achieve a SNR of 5 as a function of the distance for imaging (a and b) and differential imaging (c and d) – see Figure 3 for the experimental parameters. In the case of differential imaging, beyond a spectral resolution of a few tens the dynamics required is independent of the spectral resolution. At a distance of 10pc, the dynamics required to achieve a SNR of 5 (dotted lines) are $1.2 \times 10^6$ and $1.3 \times 10^7$ for telescope diameters of 100m and 30m, respectively.

For broad bands, the loss of SNR is only 5% for a field of view per pixel $\omega = 0.5\frac{\lambda}{D}$. $Q_p/Q_s$ is maximum in the limit of a zero field, and can be approximated within a few % precision by $(1 - e^{-p})^{-0.5}$, with $p = \lambda^2/(\gamma\delta\lambda D)$. To optimize the SNR, we will use this maximum value and the optimum value $Q_p/\omega = 0.66D/\lambda$ (Naylor 1998). Then, introducing the dynamics of the image $\mathcal{D}(\gamma)$ at the position of the planet, the SNR becomes:

$$\frac{P}{\sigma(P)} \approx \frac{f\mathcal{S}K\mathcal{D}(\gamma)}{\sqrt{8K^2(1 - e^{-p})\frac{\tau}{T} + 3.6K\mathcal{D}(\gamma)}} \quad (9)$$

At high Strehl ratios, the SNR is essentially controlled by the number of photoelectrons $K$ and by the dynamics $\mathcal{D}(\gamma)$ of the image. In the speckle noise regime, the SNR is proportional to $D^{2.5}$.

The parameter $p$ can be written as $p = \mathcal{R}/\gamma_r$, where $\gamma_r$ is the angular position of the planet expressed in Airy disk units. For broad band images, $p$ is usually smaller than 1, and the quantity $1 - e^{-p}$ can be approximated by $\mathcal{R}/\gamma_r$. Hence, the flux $F_\star$ of the star – $K = tF_\star(h\nu)^{-1}\frac{\pi}{4}D^2\delta\nu T$ – for which the speckle noise is equal to the photon noise is given by:

$$F_\star(\text{Jy}) = \frac{0.038}{t}\left(\frac{100}{D}\right)^2\left(\frac{\mathcal{D}(\gamma)}{10^6}\right)\left(\frac{\gamma_r}{100}\right)\left(\frac{0.01}{\tau}\right) \quad (10)$$

where $t$ is the overall transmission.

## 5. Differential Imaging

Differential imaging consists in making the difference between two coronagraphic images at different wavelengths in order to decrease the speckle noise at levels lower than the photon noise. We compute the difference between two images at close wavelengths $\overline{\lambda}_1$ and $\overline{\lambda}_2$, scaled at the wavelength $\lambda$ and weighted with factors $w_1 = (\overline{\lambda}_1/\lambda)^4$ and $w_2 = (\overline{\lambda}_2/\lambda)^4$ to minimize the difference between the two speckle patterns. At high Strehl ratios, this difference is given by $\Delta I = 2(1 - \mathcal{S})\Delta\lambda/\lambda < I(\alpha, \lambda >$, where $\Delta\lambda = \overline{\lambda}_1 - \overline{\lambda}_2$ is the distance between the two bands (Marois et al. 2000). Better weights could be found if we knew the exact Strehl ratio – see Eq. (3) – but it is *a priori* not the case, and in practice they have to be estimated experimentally. Assuming for simplicity the same spectral resolution $\mathcal{R} = \lambda/\delta\lambda$ for the two bands, the measured differential signal $\Delta P$ from the planet, obtained as above by adding the two symmetrical contributions, is $\Delta P = \epsilon f\mathcal{S}KQ_p$, where $\epsilon$ is the planet relative flux fraction between the two bands. Let $\sigma_{ds}$ be the differential speckle noise from the stellar halo, the variance of $\Delta P$ is given by $\sigma^2(\Delta P) = 4\sigma_{ds}^2 + 4\sigma_p^2$, with:

$$\sigma_{ds}^2 = 2K^2(1 - \mathcal{S})^2 < L(\gamma, \lambda) >^2 \left(\frac{4\lambda^2}{\pi D^2}\right)^2 Q_{ds}^2(\overline{\lambda}_1, \overline{\lambda}_2)\frac{\tau}{T} \quad (11)$$

and $Q_{ds}^2(\overline{\lambda}_1, \overline{\lambda}_2) = Q_s^2(\overline{\lambda}_1, \overline{\lambda}_1) + Q_s^2(\overline{\lambda}_2, \overline{\lambda}_2) - 2Q_s^2(\overline{\lambda}_1, \overline{\lambda}_2)$. In the limit of a zero field of view, the ratio $Q_p/Q_{ds}$ is roughly independent of $\gamma$ and can be approximated within a factor of 2 by $0.5\lambda/(1 - \mathcal{S})^{-1}\Delta\lambda^{-1}$. Hence, the SNR becomes:

$$\frac{\Delta P}{\sigma(\Delta P)} \approx \frac{\epsilon f\mathcal{S}K\mathcal{D}(\gamma)}{\sqrt{32\left[K(1 - \mathcal{S})\left(\frac{\Delta\lambda}{\lambda}\right)\right]^2\frac{\tau}{T} + 7.2K\mathcal{D}(\gamma)}} \quad (12)$$

The flux $F_\star$ of the star for which the differential speckle noise is equal to the photon noise is:

$$F_\star(\text{Jy}) = \frac{190/t}{n^2(1-\mathcal{S})^2}\left(\frac{\mathcal{R}}{100}\right)^3\left(\frac{100}{D}\right)^2\left(\frac{\mathcal{D}(\gamma)}{10^6}\right)\left(\frac{0.01}{\tau}\right) \quad (13)$$

where $n = \frac{\Delta\lambda}{\delta\lambda}$ is the distance between the two bands in resolution element units. Note that it should be possible to decrease the photon noise by a factor $\sqrt{2}$, with the use of a reference channel built by properly summing up well suited individual channels.

The scaling of the images allows optimizing the relative flux fraction $\epsilon$. Indeed after scaling, the planet contributions in the image difference are shifted apart from each other by the quantity $\gamma\Delta\lambda/\lambda = n\gamma/\mathcal{R}$. If $n\gamma/\mathcal{R} > \lambda/D$, or in other words, if $n > \mathcal{R}/\gamma_r$, they are fully separated and $\epsilon = 1$. Optimum observing conditions are reached if, for any spectral channel, it is possible to find at least another spectral channel at a distance $n$, large enough for the previous condition to hold, and small enough for the differential speckle noise to remain smaller than the photon noise. This is the case of an Earth at a distance larger than a few parsec if the total optical bandwidth is large enough. Under these conditions, the photon noise dominates and the SNR per spectral channel can be written as:

$$\frac{\Delta P}{\sigma(\Delta P)} \approx 0.4\mathcal{S}\left(\frac{f}{1.5\times 10^{-10}}\right)\left(\frac{D}{100}\right) \sqrt{tF_\star(\text{Jy})\left(\frac{100}{\mathcal{R}}\right)\left(\frac{\mathcal{D}(\gamma)}{10^6}\right)\left(\frac{T}{12}\right)} \quad (14)$$

where the integration time $T$ is expressed in hours. In the photon noise regime, the SNR on the total flux is independent of the spectral resolution. It is obtained by multiplying the SNR per spectral channel given by Eq. (14), by the square root of the number of channels. Both SNR's are proportional to $D^2$.

## 6. Discussion

We investigate the possibility to detect planets like the Earth, orbiting a star like the Sun. For this purpose we adopt a constant Earth to Sun flux ratio of $1.5\times 10^{-10}$ and Sun fluxes at 10pc of 45, 54, 55, 54, and 47Jy in the V, R, I, J and H photometric bands, respectively (Allen 1977, Wamstecker 1980, Des Marais et al. 2002). The SNR's on the planet flux are plotted in Figure 3 as a function of the distance, for an integration time of 12 hours and telescope diameters of 30 and 100m. Figures 3a and 3b represent the SNR for direct imaging with broad band images ($\lambda/\Delta\lambda = 5$), and Figures 3c and 3d represent the SNR for full contrast ($\epsilon = 1$) differential imaging, with a spectral resolution of 100 and the combined signal of 20 couples of adjacent bands. The dynamics $\mathcal{D}(\gamma)$ required to reach a SNR of 5 is plotted in Figure 4 as a function of the distance, for each observing condition. Within a factor smaller than 2, all wavelengths provide the same SNR. In the following, we discuss the detection of an Earth at a distance of 10pc.

With direct imaging, the speckle noise dominates and shorter wavelengths (V, R) provide slightly better results. A 30m telescope would give a SNR of about 0.4, which would not allow to detect the Earth. However, a 100m telescope would detect the Earth with a SNR of 8. Differential imaging provide better SNR's. In this case the photon noise dominates and the I, J and H bands are slightly more favourable. The SNR's achieved are 3 and 30 for telescopes diameters of 30 and 100m, respectively, allowing a marginal and a clear detection. For a 30m telescope, achieving a SNR of 5 would require increasing the dynamics by a factor 3 at $1.65\mu m$ to reach $1.3\times 10^7$ ($3\times 10^7$ at $0.55\mu m$). For a 100m telescope, the dynamics required for a positive detection would be 10 times smaller, i.e. $1.2\times 10^6$ at $1.65\mu m$ ($2.9\times 10^6$ at $0.55\mu m$). At the present time the best adaptive systems provide at most a dynamics at $1.65\mu m$ of $4\times 10^3$ at $0.''5$ radius at (Codona & Angel 2004). Clever methods have been proposed to reduce the residual speckle noise (Codona & Angel 2004, Guyon 2004; Labeyrie 2004) but their real performances have to be evaluated. Hence, to achieve a SNR of 5 on the flux of an Earth-like planet at a distance of 10pc with a 100m (resp. 30m) telescope, the first challenge is to produce images with a dynamics at $0.''1$ radius a few hundred times larger (resp. a few thousand) than has been presently demonstrated at $0.''5$ radius. Achieving a SNR of 5 per spectral band at a resolution of 100 would require a dynamics 20 times larger (see Figure 5).

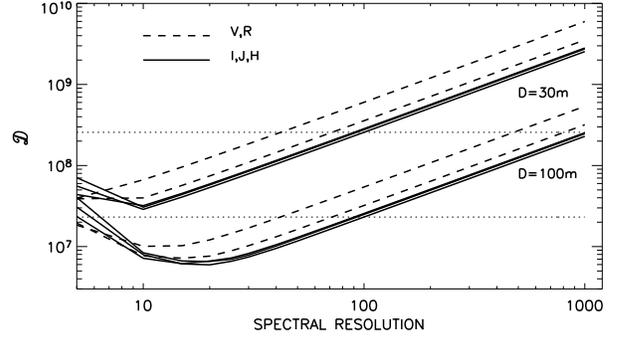

**Fig. 5.** Image dynamics required to achieve at 10pc a SNR of 5 per spectral channel as a function of the spectral resolution (differential imaging, see Figure 3 for the experimental parameters). For a spectral resolution of 100, the dynamics required (dotted lines) are $2.4\times 10^7$ and $2.6\times 10^8$ for telescope diameters of 100m and 30m, respectively. The points at a spectral resolution of 5 correspond to the dynamics required with broad band imaging.

Up to now, we have assumed a perfect optical system. In the case of real systems, especially with segmented pupils, instrumental aberrations will induce a noise similar to the speckle noise, given by:

$$\sigma^2 \approx 2K^2\sigma_a^4 <L_a(\gamma,\lambda)>^2 \left(\frac{4\lambda^2}{\pi D^2}\right)^2 Q^2 \quad (15)$$

where $Q = Q_s(\overline{\lambda}_1,\overline{\lambda}_1)$ or $Q = Q_{ds}(\overline{\lambda}_1,\overline{\lambda}_2)$. $<L_a(\gamma,\lambda)>$ and $\sigma_a$ are the incoherent extended halo and the standard deviation of the phase error, both due to instrumental aberrations. For the detection of the Earth to be effective, we need to maintain the noise induced by instrumental aberrations, below the residual speckle noise after 12 hours integration time. Hence, assuming for simplicity $<L(\gamma,\lambda)>=<L_a(\gamma,\lambda)>$, the second challenge is to control instrumental aberrations with a precision better than $\frac{\lambda}{2\pi}\left(\frac{\tau}{T}\right)^{1/4}\sqrt{1-S} \approx 1$ nanometer rms (see also Lardiere et al. 2004).

*Acknowledgements.* The author would like to thank the referee Dr. R. Racine for his helpful comments that deeply improved the clarity of the paper, and Drs M. Swain, E. Tatulli, C. Verinaud, H. Zinnecker and the Opticon ELT working group on exoplanets for helpful discussions.

## Appendix A: Long exposure coronagraphic image

The long exposure image of the telescope, at the angular position $\alpha$ and at the wavelength $\lambda$, partially corrected by adaptive optics, is given by:

$$<I_t(\alpha,\lambda)> = \frac{K}{s}\int P(\mathbf{u})P(\mathbf{v}) <\psi_i(\mathbf{u})\psi_i^*(\mathbf{v})> e^{-2i\pi\alpha.(\mathbf{u}-\mathbf{v})} d\mathbf{u}d\mathbf{v} \quad (A.1)$$

where $K$ is the average number of photoelectrons, $s = \pi D^2/(4\lambda^2)$ is the surface of the telescope in $\lambda^2$ units, $\psi_i(\mathbf{u})$ is the complex amplitude of the incoming wavefront over the telescope pupil $P(\mathbf{u})$ and, $\mathbf{u}$ and $\mathbf{v}$ are coordinates in the pupil plane in $\lambda$ units. Assuming a Gaussian statistics for the phase of the incoming wavefront, the covariance of the complex amplitude can be written as $<\psi_i(\mathbf{u})\psi_i^*(\mathbf{v})> = e^{-\frac{1}{2}\mathcal{D}(\mathbf{u};\mathbf{v})}$, where $\mathcal{D}(\mathbf{u};\mathbf{v})$ is the phase structure function. Taking into account that at high Strehl ratios $\mathcal{S}$:

$$e^{-\frac{1}{2}\mathcal{D}(\mathbf{u};\mathbf{v})} = \mathcal{S} + (1-\mathcal{S})B(\mathbf{u};\mathbf{v}) \quad (A.2)$$

we get:

$$<I_t(\alpha,\lambda)> = K\left[\mathcal{S}\frac{\pi D^2}{4\lambda^2}A(\alpha) + (1-\mathcal{S})<L(\alpha,\lambda)>\right] \quad (A.3)$$

where $A(\alpha)$ is the Airy disk function normalized at the origin and:

$$<L(\alpha,\lambda)> = \frac{1}{s}\int P(\mathbf{u})P(\mathbf{v})B(\mathbf{u};\mathbf{v})e^{-2i\pi\alpha.(\mathbf{u}-\mathbf{v})} d\mathbf{u}d\mathbf{v} \quad (A.4)$$

is the incoherent part of the image normalized in energy.

The effect of the rotation shearing coronagraph is to split the pupil into two parts, to rotate one of the pupils by 180°, to shift the phase of the other pupil by $\pi$, and then to add the two pupils. The complex amplitude $\psi_o(\mathbf{u})$ in the exit pupil at the output of the coronagraph is given by: ;

$$\psi_o(\mathbf{u}) = \frac{1}{\sqrt{2}}\left[\psi_i(\mathbf{u})P(\mathbf{u}) - \psi_i(-\mathbf{u})P(-\mathbf{u})\right] \quad (A.5)$$

Hence, the image intensity can be written as:

$$<I_c(\alpha,\lambda)> = \frac{K}{2s}\int P(\mathbf{u})P(\mathbf{v}) <\psi_i(\mathbf{u})\psi_i^*(\mathbf{v})> C(\mathbf{u},\mathbf{v},\alpha) d\mathbf{u}d\mathbf{v} \quad (A.6)$$

with

$$C(\mathbf{u},\mathbf{v},\alpha) = e^{-2i\pi\alpha.(\mathbf{u}-\mathbf{v})} - e^{-2i\pi\alpha.(\mathbf{u}+\mathbf{v})} + e^{2i\pi\alpha.(\mathbf{u}-\mathbf{v})} - e^{2i\pi\alpha.(\mathbf{u}+\mathbf{v})} \quad (A.7)$$

After some transformations, we get:

$$<I_c(\alpha,\lambda)> = K(1-\mathcal{S})\left[<L(\alpha,\lambda)> - <L_+(\alpha,\lambda)>\right] \quad (A.8)$$

with:

$$<L_+(\alpha,\lambda)> = \frac{1}{s}\int P(\mathbf{u})P(\mathbf{v})B(\mathbf{u};\mathbf{v})e^{-2i\pi\alpha.(\mathbf{u}+\mathbf{v})} d\mathbf{u}d\mathbf{v} \quad (A.9)$$

For a stationnary structure function, $B(\mathbf{u};\mathbf{v}) = B(\mathbf{u}-\mathbf{v})$, and a pupil much larger than the width of $B(\mathbf{u}-\mathbf{v})$, $L_+(\alpha,\lambda) \approx L(\alpha,\lambda)A(2\alpha)$. Hence, the long exposure coronagraphic image can be written as:

$$<I_c(\alpha,\lambda)> \approx K(1-\mathcal{S})<L(\alpha,\lambda)>[1-A(2\alpha)] \quad (A.10)$$

The effect of the coronagraph is to cancel the coherent part of the light. In the case of a system formed by an unresolved star in the center and a planet at a distance $\gamma$, the image of the planet appears as two symmetrical contributions superposed on the stellar halo.

## Appendix B: Covariance of speckle noise

The spatial and spectral covariance of the speckle noise is given by:

$$C(\alpha,\lambda_1;\beta,\lambda_2) = K^2 \frac{l_1^2 l_2^2}{4s^2} \int d\mathbf{u} d\mathbf{v} d\mathbf{u}' d\mathbf{v}'$$
$$P(\mathbf{u})P(\mathbf{v})P(\mathbf{u}')P(\mathbf{v}')C(\mathbf{u},\mathbf{v},\alpha l_1)C(\mathbf{u}',\mathbf{v}',\beta l_2) \quad (B.1)$$
$$[<\psi_i(\mathbf{u},\lambda_1)\psi_i^*(\mathbf{v},\lambda_1)\psi_i(\mathbf{u}',\lambda_2)\psi_i^*(\mathbf{v}',\lambda_2)> -$$
$$<\psi_i(\mathbf{u},\lambda_1)\psi_i^*(\mathbf{v},\lambda_1)><\psi_i(\mathbf{u}',\lambda_2)\psi_i^*(\mathbf{v}',\lambda_2)>]$$

where $l_1 = \lambda/\lambda_1$ and $l_2 = \lambda/\lambda_2$. In first approximation, optical pathes are achromatic and $\phi_i(\mathbf{u},\lambda_1) = l_1\phi_i(\mathbf{u},\lambda)$. Then developping the fourth order moment of the complex amplitude following the method of Korff (1973), we get:

$$C(\alpha,\lambda_1;\beta,\lambda_2) = K^2 \frac{l_1^2 l_2^2}{4s^2} \int d\mathbf{u} d\mathbf{v} d\mathbf{u}' d\mathbf{v}'$$
$$P(\mathbf{u})P(\mathbf{v})P(\mathbf{u}')P(\mathbf{v}')C(\mathbf{u},\mathbf{v},\alpha l_1)C(\mathbf{u}',\mathbf{v}',\beta l_2)e^{-\frac{1}{2}l_1^2 \mathcal{D}(\mathbf{u};\mathbf{v})} \quad (B.2)$$
$$e^{-\frac{1}{2}l_2^2 \mathcal{D}(\mathbf{u}';\mathbf{v}')}\left[e^{-\frac{1}{2}l_1 l_2[\mathcal{D}(\mathbf{u};\mathbf{v}')+\mathcal{D}(\mathbf{v};\mathbf{u}')-\mathcal{D}(\mathbf{u};\mathbf{u}')-\mathcal{D}(\mathbf{v};\mathbf{v}')]} - 1\right]$$

We assume high Strehl ratio ($\mathcal{S} > 0.5$) and we introduce Eq. (A.2) into Eq (B.2). Then, we make a series expansion to the second order with respect to $\frac{1-\mathcal{S}}{\mathcal{S}}$ of the expression between brackets within the integral. Making the approximation: $e^{-\frac{1}{2}[l_1^2 \mathcal{D}(\mathbf{u};\mathbf{v})+l_2^2 \mathcal{D}(\mathbf{u}';\mathbf{v}')]} \approx \mathcal{S}^{l_1^2+l_2^2}$, and retaining the dominant terms, we get:

$$C(\alpha,\lambda_1;\beta,\lambda_2) \approx 2K^2 l_1^4 l_2^4 \mathcal{S}^{l_1^2+l_2^2-2}(1-\mathcal{S})^2$$
$$\left(\frac{1}{s}\int P(\mathbf{u})P(\mathbf{v})B(\mathbf{u};\mathbf{v})e^{-2i\pi(\alpha l_1 \mathbf{u}-\beta l_2 \mathbf{v})}d\mathbf{u}d\mathbf{v}\right)^2 \quad (B.3)$$

Assuming a stationnary structure function and a large pupil, the covariance coefficient can be written as:

$$C(\alpha,\lambda_1;\beta,\lambda_2) \approx 2 < I(\alpha,\lambda)>^2 l_1^4 l_2^4 \mathcal{S}^{l_1^2+l_2^2-2} A(\alpha l_1 - \beta l_2)$$
$$\approx 2 < I(\alpha,\lambda_1)>< I(\beta,\lambda_2)> A(\alpha l_1 - \beta l_2) \quad (B.4)$$